\documentclass[prd,english,aps,showpacs,nofootinbib,preprint,eqsecnum]{revtex4-1}
\usepackage{graphicx,color,amsmath,amsxtra}
\usepackage{epsf}
\usepackage{amssymb}
\usepackage{enumerate}
\usepackage{hhline}
\usepackage{array}
\usepackage{tabularx}
\newcommand{\beq}{\begin{equation}}
\newcommand{\eeq}{\end{equation}}
\newcommand{\beqa}{\begin{eqnarray}}
\newcommand{\eeqa}{\end{eqnarray}}
\newcommand{\barr}{\begin{array}}
\newcommand{\earr}{\end{array}}
\newcommand{\bib}[1]{\bibitem{#1}}
\begin{document}

\title{Matter Density Perturbations in Modified Teleparallel Theories}
%%% of Dark Energy}
\date{\today}

\author{Yi-Peng Wu$^{1}$\footnote{E-mail address: s9822508@m98.nthu.edu.tw} 
and
Chao-Qiang Geng$^{1,2,3}$\footnote{E-mail address: geng@phys.nthu.edu.tw}     
}
\affiliation{
$^1$Department of Physics, National Tsing Hua University, Hsinchu, Taiwan 300 
\\ 
$^2$Physics Division, National Center for Theoretical Sciences, Hsinchu, Taiwan 300
\\
$^3$College of Mathematics \& Physics, Chongqing University of Posts \& Telecommunications, Chongqing, 400065, China
}
%==================================================================================================================================%
%--------------------------------------Abstract----------------------------------------------------------------------------------%
\begin{abstract}

We study the matter density perturbations
in modified teleparallel gravity theories, where extra
degrees of freedom arise from the local Lorentz violation
in the tangent space. We formulate a vierbein perturbation
with variables addressing all the 16 components of the 
vierbein field. By assuming the perfect fluid matter source,
we examine the cosmological implication of the 6 unfamiliar 
new degrees of freedom in modified $f(T)$ gravity theories. 
We find that despite the new modes in the vierbein scenario 
provide no explicit significant effect in the small-scale regime, 
they exhibit some deviation from the standard 
general relativity results in super-horizon scales.  

\end{abstract}


%\pacs{}
\maketitle
%==================================================================================================================================%
%----------------------------------------Section I. Introduction-------------------------------------------------------------------%
\section{Introduction}

%The late time cosmic acceleration can be explained by scalar-tensor and $f(R)$ theories.

Theories constructed in ``Teleparallelism'' have been widely considered 
as an alternative origin to explain the acceleration of the cosmic
expansion of the Universe. These models are mainly achieved by modifying the gravitational Lagrangian
of the teleparallel equivalence of general relativity (TEGR)  
which reveals an equivalent formulation of classical gravity 
from general relativity~\cite{E,HT}. 
It has seen modifications in the favor of  the
non-linear generalization of TEGR, 
known as $f(T)$ gravity theories~\cite{ft}, 
or by introducing a scalar with the non-minimal 
coupling to the gravity~\cite{TDE}.

However, perhaps the most important feature for those 
modified teleparallel theories~\cite{ft,TDE} is the introduction of 
extra degrees of freedom (EDoFs) due to the lack of 
local Lorentz invariance. These EDoFs are unfamiliar to 
the usual metric scenario since they merely contribute as the total 
divergence in the simplest teleparallel construction, i.e. TEGR.  
To be more precisely, there are 6 components of the vierbein 
field released from the gauge freedom of the local Lorentz transformation in the 
tangent frame, which become physical modes to be determined by
the field equations~\cite{LV}. In a further analysis of 
$f(T)$ theories~\cite{dof}, 
it has been found that modified teleparallel theories contain three
additional physical degrees of freedom (ADoFs) 
to the usual massless spin-2 graviton.
Via conformal transformations~\cite{CT} or
scalar-tensor formulations of the $f(T)$ theories~\cite{GTG}, 
one of the ADoFs reveals explicitly as the (conformal) 
scalar mode, which is related to the derivative
of the arbitrary function $f$. This scalar degree of freedom 
is also familiar in the ad hoc study of $f(R)$ theories, 
thus does not belong to EDoFs. 
The result from the counting of degrees of freedom 
indicates that only two of the 6
EDoFs  will turn into physical modes with 
dynamical importance beyond the metric perturbation scenario.

As discussed in~\cite{LSS}, from a covariant and gauge invariant 
approach of  $f(T)$ gravity, 
these unfamiliar physical quantities can have virtual significant effects
during the cosmological evolutions. In the present work, 
we study the matter density perturbations of  modified teleparallel 
theories with regarding to all the dynamical 
variables of the vierbein field.
First of all, we illustrate perturbation modes of the
vierbein field including the usual variables well studied in the 
metric perturbations as well as the unfamiliar components induced from 
the lack of Lorentz symmetry. Our formulation of the perturbed 
variables is able to separate the analysis into scalar, vector 
and tensor modes, as the similar scenario in the metric perturbations.
Moreover, the 6 EDoFs in \textit{vierbein perturbations} appear 
to be described by a spatial vector and a dynamical 
spatial antisymmetric tensor with each containing 3 independent components.

We specify our study to the matter density perturbations 
in $f(T)$ gravity by examining the behavior of
the EDoFs. Under the perfect fluid assumption of
the matter source, we find that the scalar modes of
EDoFs  perform with dynamical significance;
while the new vector modes remain decaying modes only.
Nevertheless, the perturbed equations indicate that
these new scalar modes have a mere implicit contribution
to the evolution of the density perturbations for
sub-horizon scales $k\gg aH$, where $k,a$ and $H$ 
are the wave number, the scale factor and Hubble parameter
respectively. In the super-horizon scales $k\ll aH$, it
is the new mode of EDoFs   that causes the deviation from
general relativity.
These results agree with the previous
finding from both metrical or non-metrical approaches
 \cite{LSS,CP of fT,fT p}.

The paper is organized as follows. 
In Sec.~II, we review modified teleparallel theories. 
In Sec.~ III, we illustrate the variables in vierbein perturbations.
 We apply our formulation to $f(T)$ gravity in Sec.~IV.
Finally, conclusions are given in Sec.~V.

%======================================================================================================%
%----------------------------------------Section II. Teleparallel gravity----------------------------------------%
\section{Modified Teleparallel Theories}

The teleparallel formalism uses the vierbein field 
${\mathbf{e}_A(x^\mu)}$ 
as the dynamical variables, which form an orthonormal basis 
of the tangent space:
$\mathbf{e}_A\cdot%
\mathbf{e}_B\equiv\eta_{AB}=diag (1,-1,-1,-1)$. 
The vector $\mathbf{e}_A$ is commonly addressed by its
components $e_A^\mu$ in a coordinate basis, that is
$\mathbf{e}_A = e^\mu_A\partial_\mu$, 
while the metric tensor is obtained from
the dual vierbein as 
$g_{\mu\nu}=\eta_{AB}\, e^A_\mu \, e^B_\nu $.\footnote{For 
the \textsl{coordinate and tangent frames}, Greek  indices $\mu, \nu,...$ and 
capital Latin indices $A,B,...$ run over space and time, while Latin indices 
$i,j,...$ and $a,b,...$, 
represent the spatial part of 1, 2, 3, respectively.}
Although the ``distance'' in teleparallel gravity is still determined
by the metric tensor $g_{\mu\nu}(x)$,
the gravitational effect is, 
instead of the concept of curvature in general relativity, 
geometrized purely by the torsion tensor, 
\beq
T^{\rho}_{\,\,\mu\nu}=e^\rho_A(\partial_\mu e^A_\nu-\partial_\nu e^A_\mu),
\eeq
which is composed by the subtraction of the curvatureless connection:
$\Gamma^{\rho}_{\,\,\mu\nu}=e^\rho_A\partial_\nu e^A_\mu$.
Such curvatureless connection defines an absolute parallel transportation of
the objects on the manifold with only regard to the torsion effect.
In order to arrive at some second order field equations, the teleparallel
gravity Lagrangian appears the quadratic of the torsion tensor,
while TEGR
is found to be of the specific choice~\cite{HT}:
\beqa
\label{TEGRL}
\mathcal{L}_{TEGR}=\frac{1}{2}T&=&
                   \frac{1}{8}T_{\rho}^{\,\,\mu\nu}T^{\rho}_{\,\,\mu\nu}-
                   \frac{1}{4}T^{\mu\nu}_{\,\,\,\,\,\,\,\,\rho} T^{\rho}_{\,\,\mu\nu}-
                   \frac{1}{2}T^{\rho}_{\,\,\rho\mu}T^{\nu\,\,\mu}_{\,\,\nu}.
\eeqa
Denoting the tensor 
$S_\rho^{\;\,\mu\nu}=\frac{1}{2}(K_{\;\;\;\rho}^{\mu\nu}+
\delta^\mu_\rho T^{\alpha\nu}_{\;\;\;\alpha}-
\delta^\nu_\rho T^{\alpha\mu}_{\;\;\;\alpha})$ 
where
$K^{\rho}_{\,\,\mu\nu}=\frac{1}{2}(T^{\,\,\rho}_{\nu\;\,\mu} +
 T^{\,\,\rho}_{\mu\;\,\nu}-T^{\rho}_{\,\,\mu\nu})$,
the variation of the TEGR action
$S=\frac{1}{2}\int d^4x e[T/\kappa^2+\mathcal{L}_m]$ 
with respect to vierbein gives the field equation
\begin{eqnarray}\label{eom}
e^{-1}\partial_{\mu}(ee_{A}^{\rho}S_{\rho}{}^{\mu\nu})
-e_{A}^{\lambda}T^{\rho}{}_{\mu\lambda}S_{\rho}{}^{\nu\mu}
-\frac{1}{4}e_{A}^{\nu}T
= \frac{\kappa^2}{2}e_{A}^{\rho}\Theta_{\rho}{}^{\nu},
\end{eqnarray}
where 
$e_{A}^{\rho}\Theta_{\rho}{}^{\nu}\equiv e^{-1}\delta\mathcal{L}_m 
/\delta e^A_\nu$ is the energy-momentum tensor of  matter.
It is noteworthy that 
\beq \label{ET}
2e^{-1}\partial_{\mu}(ee_{A}^{\rho}S_{\rho}{}^{\mu\nu})
 -2e_{A}^{\lambda}T^{\rho}{}_{\mu\lambda}S_{\rho}{}^{\nu\mu}
 -\frac{1}{2}e_{A}^{\nu}T=G_{A}^\nu,
\eeq  
is nothing but an equivalent mathematical manipulation of
the Einstein tensor, 
thus  (\ref{eom}) illustrates identically the geometrical 
formulation of general relativity.

The equivalence formulation TEGR has received many extensions for 
the cosmological purpose of the late time accelerating universe.
One of the common modification is inspired from $f(R)$ gravity
to generalize the torsion scalar to become an 
arbitrary function $T \rightarrow f(T)$.
For the interest of dark energy phenomena, it is usually to consider
the action of the form
\begin{equation}
S=\frac{1}{2\kappa^{2}}\int\,d^{4}x e\left[T + f(T)
+\mathcal{L}_m\right], \label{action2}
\end{equation}
where $e=\mbox{det}(e^A_\mu=\sqrt{-g})$. The action (\ref{action2})
provides the field equation
\begin{eqnarray}
%\label{eom2}
\left(1+f_T \right)
\left[e^{-1}e^A_\mu\partial_{\lambda}(ee_A^{\rho}S_{\rho}{}^{\lambda\nu} )
      -T^{\rho}{}_{\lambda\mu}S_{\rho}{}^{\nu\lambda}\right]
      -\frac{1}{4}\delta_{\mu}^{\nu} \left[T+f(T) \right]\nonumber\\ 
 + S_{\mu}{}^{\lambda\nu}\left(\partial_\lambda  f_T\right)
=\frac{\kappa^2}{2}\Theta_{\mu}{}^{\nu},~~~~~~~~~~~~
\label{eom2}
\end{eqnarray}
which is obtained by variation with respect to the vierbein
and then transit the tangent frame indices to coordinate ones.
We use the notation $f_T\equiv\partial f(T)/\partial T$
and $f_{TT}\equiv\partial^2f(T)/\partial T^2$ and so on.
The critical issue in teleparallel gravity arises from
the fact that the curvatureless connection
$\Gamma^{\rho}_{\,\,\mu\nu}=e^\rho_A\partial_\nu e^A_\mu$
is not an invariant quantity under the local Lorentz
transformation in the tangent frame
$e^A{}_\mu=\Lambda^A{}_B(x)e^B{}_\mu$, where
$\eta_{CD}=\Lambda^A_C\Lambda^B_D\eta_{AB}$.
Hence, both $T^{\rho}_{\,\,\mu\nu}$ and $T$ are
Lorentz violation quantities as well~\cite{T intro}.
It can be seen that
the left hand side of  (\ref{eom2}) shows no 
symmetric property between the two indices $\mu$ and $\nu$;
while, on the other hand, the energy-momentum tensor 
$\Theta_{\mu\nu}$ has to be symmetric
due to the invariance principle of the local Lorentz transformation
in the matter sector 
(see verifications in both \cite{LV} and \cite{Weinberg}).
By using the relation (\ref{ET}), the field equations (\ref{eom2})
can be rewritten into the covariant version as
\beq 
\left( 1+f_T\right)G_\mu{}^\nu-\frac{1}{2}\delta_\mu{}^\nu(f-Tf_T)
+2 S_{\mu}{}^{\lambda\nu}\left(\partial_\lambda  f_T\right)
=\kappa^2 \Theta_\mu{}^\nu.
\eeq
It becomes evident from this version that the antisymmetrization of 
the field equations leads to some non-trivial constraints of the vierbein:
\beq \label{constraint} 
\left(g^{\mu\alpha}S_{\mu}{}^{\lambda\beta}-
      g^{\nu\beta} S_{\nu}{}^{\lambda\alpha}\right)\partial_\lambda f_T=0. 
\eeq 
These 6 additional equations imply the existence of
the 6 e.ds.o.f as the consequence of the lack of 
 local Lorentz symmetry in the teleparallel formalism.
Nevertheless, in the TEGR limit  of $f_T\rightarrow const.$ Eq.
(\ref{constraint}) automatically vanishes and the dynamical
degrees of freedom in teleparallel gravity reduce to be the same
as general relativity.

It is remarkable in the context of the dark energy interest
that  (\ref{constraint}) disappears identically 
in the background vierbein choice for the flat FRW geometry:
\beq \label{background v}
e^A_{\mu}=\mbox{diag}(1,a,a,a).
\eeq
In the perturbation study of modified teleparallel theories,
however, such additional constraints are virtually important 
as they govern the equations of motion for those
new dynamical degrees of freedom.
 
%======================================================================================================%
%----------------------------------------Section II. Linear Perturbations in Teleparallel Formulations----------------------------------------%
\section{Dynamical Variables in Vierbein Perturbations}

Although the metric perturbation is conventionally addressed by
variables corresponding to the 10 degrees of freedom of
the metric tensor, perturbations in teleparallel gravity,
however, demand to find variables for all the 16 components of
the vierbein field. It is noticeable that the
part of the variables, depicting the 6 EDoFs, shall
show no contribution in the specific limit of TEGR;
while the metric scenario must be recovered by the
other 10 degrees of freedom. 
This requirement can be seen by a decomposition
of the vierbein as
\beq \label{decomposition}
e^A_\mu(x)=\bar{e}^A_\mu(x)+\ae^A_\mu(x),
\eeq
which satisfies the condition
\beq \label{condition}
g_{\mu\nu}(x)=\eta_{AB}\, e^A_\mu (x)\, e^B_\nu (x)
             =\eta_{AB}\, \bar{e}^A_\mu(x)\,\bar{e}^B_\nu(x),
\eeq
where $\bar{e}^A_\mu$ illustrates the part of vierbein 
quantities that are familiar in the metric perturbations. 
In fact, the perturbed variables of
$\bar{e}^A_\mu$ have been considered in \cite{CP of fT} of the form
\begin{eqnarray} 
\label{p vierbein}
\bar{e}^0_\mu &=&\delta^0_\mu(1+\psi)+a\delta^i_\mu(G_i+\partial_iF), \\
\bar{e}^a_\mu &=&a\delta^a_\mu(1-\varphi)+
        a\delta^i_\mu(h^a_i+\partial_i\partial^a B +\partial^a C_i), 
        \nonumber
\end{eqnarray}
which give rise to the usual perturbed metric
\begin{eqnarray} \label{p-metric}
g_{00} &=& 1+2\psi, \\ \nonumber
g_{i0} &=& a(\partial_iF + G_i), \\ \nonumber
g_{ij} &=& -a^2[(1-2\varphi)\delta_{ij}+h_{ij}+\partial_i\partial_jB
           +\partial_jC_i+\partial_iC_j], ~~~~~~~
\end{eqnarray}
with scalar modes $\varphi$ and $\psi$, transverse vector modes $C_i$
and $G_i$ as well as the transverse traceless tensor mode $h^a_i$. 
Note that our notations are used only $\delta_{ab}$ to the upper and lower  
spatial indices with $\eta_{ab}\equiv-\delta_{ab}$, and also to the transition
between frames $h_{ij}=\delta_{ai}h^a_j$.

It is evident from the decomposition (\ref{decomposition}) 
that all the unfamiliar part of the vierbein,
$\ae^A_\mu$, will not appear in any metrical
quantities such as the Ricci scalar $R$ or the Einstein tensor $G_{\mu\nu}$. 
The specific formulation of teleparallel gravity, TEGR, involves no
contribution from $\ae^A_\mu$ as well, given that the torsion
scalar $T$  differs from $R$ by a divergent term.
Moreover, in the context following the flat FRW background
$e^A_{\mu}=\mbox{diag}(1,a,a,a)$, $\ae^A_\mu$
reveals as purely perturbed quantities. The contribution of this unfamiliar
component, thus, appears only to the perturbation equations of modified
teleparallel theories.

In order to present a further
study, we denote each part of $\ae^A_\mu$ as
\beqa  
\ae^0_\mu &=&\delta^0_\mu\ae+\delta^i_\mu\ae_i, \\
\ae^a_\mu &=&\delta^0_\mu A^a+\delta^i_\mu B^a_i. \nonumber
\eeqa
These components are not independent from each other
as, to the linear order, the condition (\ref{condition}) leads to 
\beq
\ae=0\;,\; \ae_i=aA_i\;,\; \mbox{and} \;B_{ij}+B_{ji}=0\;. 
\eeq
As a result, the dynamical variables of
$\ae^A_\mu$ are described by a vector $A^i$ and a spatial 
antisymmetric tensor $B_{ij}$, which contain overall $3+3=6$ 
degrees of freedom.
These degrees of freedom address completely of those of EDoFs released
from the local Lorentz violation in the teleparallel formalism. 
A similar decomposition analogous to the metric variables is 
also treated to these new quantities as\footnote{There is no reference for the decomposition of an antisymmetric
tensor in cosmological perturbations. We treat $\partial_jB_{ji}$ as
a spatial vector for that $B_{ij}$ only presents in the perturbed
(\ref{eom2}) of this form, see the discussion in Sec. IV.}
\beq \label{e.d.o.f}
A^i=\partial^i \alpha +\alpha^i\;;\;
\partial_jB^{ji}\equiv\partial^i\beta +\beta^i,~~
\eeq
where $\alpha^i$ and $\beta^i$
are transverse vectors, which satisfy
$\partial_i\alpha^i=\partial_i\beta^i=0$,
and hence,
$\partial_i\partial_jB^{ji}=\partial^2\beta=0$.

In the perturbation theory for modified teleparallel gravity models,
both variables of $\bar{e}^A_\mu$ and $\ae^A_\mu$ 
dynamically contribute to the cosmological evolution.
Namely, in the following we will consider  the scalar part
of the perturbed vierbein
\beqa \label{gauge-ready scalar mode}
e^0_\mu &=&\delta^0_\mu(1+\psi)+a\delta^i_\mu\partial_i(F+\alpha), \\
e^a_\mu &=&a\delta^a_\mu(1-\varphi)+a\delta^i_\mu(\partial_i\partial^a B
         +B^a{}_i)+ \delta^0_\mu\partial^a\alpha,\nonumber
\eeqa
as well as the vector one
\begin{eqnarray}
 \label{gauge-ready vector mode}
e^0_\mu &=&\delta^0_\mu+a\delta^i_\mu(G_i+\alpha_i), \\
e^a_\mu &=&a\delta^a_\mu+a\delta^i_\mu(\partial^a C_i+B^a{}_i
    )+ \delta^0_\mu\alpha^a. \nonumber ~~~~~~~
\end{eqnarray}
We  remark that the 16 degrees of freedom 
in the \textsl{vierbein perturbation} are composed
separately by 6 scalar modes:
$\psi$, $\phi$, $B$, $F$, $\alpha$ and $\beta$; and 
4 vector modes: $C_i$, $G_i$, $\alpha_i$ and $\beta_i$;
as well as a transverse traceless tensor mode $h_{ij}$.

Before proceeding the calculations to the theories of our concern,
we shall review some gauge issues similar to the scenario of metric 
perturbations.
Even though teleparallel gravity is formulated via the vierbein with
an explicit reference to the tangent frame, the theories are still
described by covariant tensors of the spacetime so that are invariant
under the general coordinate transformations 
$x^\mu\rightarrow x^\mu+\epsilon^\mu(x)$.
This transformation changes the vierbein field by 
$\delta e_\mu^A = -\bar{e}_\lambda^A\partial_\mu \epsilon^\lambda-
 \epsilon^\lambda\partial_\lambda\bar{e}_\mu^A$
to the linear order, and provides some gauge choices for us 
to eliminate some part of the perturbed variables. 
To proceed any gauge choice,
$\epsilon^\mu$ is separately treated by its temporal part $\epsilon_0$
and spatial vector one $\epsilon^i$,
while the spatial part can be decomposed into a spatial scalar plus a
transverse vector: $\epsilon_i=\partial_i\epsilon^S+\epsilon^V_i\,,\;
\partial_i\epsilon^V_i=0$. The varied vierbein $\delta e_\mu^A$
then gives
\beqa \label{gauge vierbein}
\delta e_\mu^0&=&-\delta^0_\mu\partial_0\epsilon^0-\delta^i_\mu\partial_i \epsilon^0 \\
\delta e_\mu^a&=&-a\delta^0_\mu(\partial_0\partial^a\epsilon^S+\partial_0\epsilon_V^a) 
  -a\delta^i_\mu(\partial^a\partial_i\epsilon^S+\partial_i\epsilon_V^a+\epsilon^0\delta^a_i),
  \nonumber
\eeqa
and this transformation changes the metric tensor by \cite{Weinberg}
\beq
\delta g_{\mu\nu}=-\bar{g}_{\lambda\mu}\partial_\nu\epsilon^\lambda
                  -\bar{g}_{\lambda\nu}\partial_\mu\epsilon^\lambda
                  -\epsilon^\lambda\partial_\lambda\bar{g}_{\mu\nu}. 
\eeq
It becomes straightforwardly to proceed any gauge choice 
by eliminating the variables in vierbein perturbations 
from  (\ref{gauge vierbein}).
For instance, the choice for the Longitudinal gauge and synchronous 
gauge has been considered in \cite{CP of fT}.

%==================================================================================================================================%
%----------------------------------------Section III. Perturbation Equations of Teleparallel Dark Energy-----------------------------------------------------------%
\section{Perturbation Equations in $f(T)$ gravity}

In this section, we study the matter density perturbations of 
$f(T)$ gravity base on the vierbein perturbations
(\ref{gauge-ready scalar mode}) and (\ref{gauge-ready vector mode}). 
The background matter source is assumed to be a perfect fluid of
the form
\beq
\Theta_{\mu\nu}=pg_{\mu\nu}-(\rho+p)u_\mu u_\nu,
\eeq 
where $u^\mu$ is the fluid 4-velocity. Using the background
choice (\ref{background v}), we have $T=-6H^2$, 
and the field equations (\ref{eom2}) then read
\beqa \label{bg}
3H^2    &=& \kappa^2\rho-\frac{f(T)}{2}-6f_T H^2 \\
2\dot{H}&=& -\frac{\kappa^2(\rho+p)}{1+f_T-12H^2f_{TT}},
\eeqa
where $H\equiv \dot{a}/a$ is the Hubble parameter.
The matter density $\rho$ includes such as pressureless 
dust-like matter $\rho_m$ and radiation $\rho_r$,
and satisfies the continuity equation
$\dot{\rho}+3H(\rho+p)=0$.
In what follows we separate our discussion into
scalar and vector parts. The results of scalar
perturbations are investigated in both sub-horizon
and super-horizon scales.

%=====================================================================================%
%----------Subsection B. Scalar Perturbations------------------------------%
\subsection{Scalar Perturbations}

We consider in the following to eliminate the variables $F$, $B$
and $C_i$ in (\ref{p-metric}) by properly choosing $\epsilon^0$, $\epsilon^S$
and $\epsilon^V_i$, respectively. Namely, we proceed the calculation
with the scalar part of the vierbein perturbations given as
\beqa \label{scalar mode}
e^0_\mu &=&\delta^0_\mu(1+\psi)+a\delta^i_\mu\partial_i\alpha \\
e^a_\mu &=&a\delta^a_\mu(1-\varphi)+a\delta^i_\mu B^a{}_i+ 
            \delta^0_\mu\partial^a\alpha.
\eeqa
This provides the metric in terms of the \textsl{Longitudinal gauge}:
\beq \label{l-g}
ds^2=(1+2\psi)dt^2-a^2(1-2\varphi)\delta_{ij}dx^idx^j,
\eeq
which is commonly used for the matter density perturbations in 
modified gravity theories. 

\subsubsection{the effective gravitational coupling}

We shall begin the discussion with 
the matter source as this involves no difference between the
metric and vierbein scenarios.
We consider the perturbed energy-momentum tensor $\Theta_\mu{}^\nu$, 
given by 
\beq
\Theta^0_0=-(\rho+\delta\rho),\; \Theta^0_i=-(\rho+p)\delta u_i,\; 
\Theta^i_j=(p+\delta p)\delta^i_j +\partial^i\partial_j\pi,
\eeq
where $\delta u_i$ characterizes the velocity perturbation of the fluid
and $\pi$ is the so-called anisotropic stress.
In the same manner, $\delta u_i$ shall be decomposed into
a scalar vector potential $\delta u$ and a transverse
vector $\delta u_i^V$.
For a pressureless matter, i.e. $p_m=0$, the standard 
continuity equation becomes
\beq
\dot{\rho}_m+3H\rho_m=0.
\eeq
The conservation of energy-momentum 
$\nabla^\mu \Theta_{\mu\nu}=0$,\footnote{This is 
conserved with respect to the metric covariant
derivative, which can be derived from the invariance under both
general coordinate and local Lorentz transformations
of the matter action.  See also~\cite{LV, Weinberg}.}
thus gives the equations
of motion in the Fourier space as \cite{gauge-ready}:
\beqa
\delta \dot{\rho}_m+3H\delta\rho_m &=&-\rho_m 
\left( 3\dot{\varphi}+\frac{k^2}{a}\delta u_m \right), \\
                     \delta \dot{u}&=&-\psi,
\eeqa
where $k$ is a co-moving wave number and $\delta u\equiv a\delta u_m$. 
It is conventional to define the gauge invariant variable
\beq
\delta_m =\frac{\delta\rho_m}{\rho_m}+3H\delta u
\eeq
so that under the sub-horizon approximation, $k\gg aH$, 
we obtain the evolution equation of $\delta_m$ in 
the Longitudinal gauge as
\beq  \label{psi eq}
\ddot{\delta}_m+2H\dot{\delta}_m-\frac{k^2}{a^2}\psi \simeq 0.
\eeq 
This expression is convenient to compare with the 
standard matter perturbation equation:
\beq  \label{G eff}
\ddot{\delta}_m+2H\dot{\delta}_m-4\pi G_{eff}\rho_m\delta_m = 0,
\eeq 
with $G_{eff}$ is the effective Newton's gravitational constant, which is equal to
$G$ in general relativity.

For the gravity sector of the vierbein perturbations in 
$f(T)$ theories, the torsion tensors from the perturbed vierbein 
(\ref{scalar mode}) become 
\beqa \nonumber
T^{0}{}_{0i}&=&-\partial_i\psi+ a\partial_0\partial_i\alpha, \\
T^{i}{}_{0j}&=&(H-\dot{\varphi})\delta^i_j+ \partial_0B^i{}_j
               -a^{-1}\partial_j\partial^i\alpha ,  \\ \nonumber
T^{i}{}_{jk}&=&\partial_k(\delta^i_j\varphi-B^i{}_j)-
               \partial_j(\delta^i_k\varphi-B^i{}_k),
\eeqa
and the torsion scalar is given by
\beq
T=-6H^2+12H(\dot{\varphi}+H\psi)+4a^{-2}\partial^2\alpha_m,
\eeq
where $\alpha_m\equiv aH\alpha$.
We can denote $T=\bar{T}+\delta T$, where $\bar{T}$ and $\delta T$ are the background and  perturbed parts, given by
 $\bar{T}=-6H^2$ and
$\delta T\equiv 12H(\dot{\varphi}+H\psi)+4a^{-2}\partial^2\alpha_m$, respectively.
As a result, the perturbations of $f$ and $f_T$ are decomposed into
$f=\bar{f}+\delta f$ and
$f_T=\bar{f}_T+\delta f_T$ with
$\delta f=f_T\delta T$ and $\delta f_T=f_{TT}\delta T$. 
The bar of the background component
will be omitted in the following discussion for simplicity.
The scalar perturbations of the field equations (\ref{eom2}) 
with the matter source $p_m=\delta p=0$ are
\beqa
(1+f_T)\left[  6H(\dot{\varphi}+H\psi)-2\frac{k^2}{a^2}\varphi\right]
-6 H^2\delta f_T 
                          &=&-\kappa^{2}\delta\rho_m 
                          \label{00}, \\
2(1+f_T) (\dot{\varphi}+H\psi)
+ 2f_{TT}\dot{T}\left( \varphi+\frac{1}{2}\beta\right) 
                          &=&-\kappa^{2}\rho_m \delta u  
                          \label{0i}, \\
-2(1+f_T) (\dot{\varphi}+H\psi)+2 H \delta f_T
                          &=&\kappa^{2}\rho_m \delta u   
                          \label{i0}, \\ \nonumber
(1+f_T)
\left[ 12H(\dot{\varphi}+H\psi)+2(H\dot{\psi}+\ddot{\varphi}+2\dot{H}\psi) \right]&&
         \\ 
+2f_{TT}\dot{T}(\dot{\varphi}+2H\psi)-3H^2 \delta f_T -2H \delta\dot{f}_T 
                          &=& 0,    \label{ij}                                             
\eeqa
where $\partial^2\equiv \delta^{ij}\partial_i\partial_j$,
while the zero anisotropic stress assumption ($\pi=0$) leads to 
\beq \label{psi=phi}
\psi=\varphi + \frac{12\dot{H}f_{TT}}{1+f_T}\alpha_m.
\eeq
Note that $\delta\dot{f}_T$ is the brief for 
$\partial_0(\delta f_T)=
\dot{f}_{TT}\delta T+f_{TT}\partial_0(\delta T)$.
The trace equation corresponding to $\Theta^\mu_\mu$ is
\beqa \label{trace}
(1&+&f_T)\left[ 24H(\dot{\varphi}+H\psi)+3\partial_0(\dot{\varphi}+H\psi)\right. 
+3\left. \dot{H}\psi+\frac{k^2}{a^2}(\psi-2\varphi+4H\alpha_m)\right] \nonumber \\
&-&3(\dot{H}+4H^2)\delta f_T-(1+f_T)\delta T  
 +3f_{TT}\dot{T}(\dot{\varphi}+2H\psi)-3H\delta\dot{f}_T
 = \frac{\kappa^2}{2}\delta\rho_m, 
\eeqa
 derived from  (\ref{00})
and (\ref{ij}).

Nonetheless, to complete the perturbations in 
$f(T)$ gravity, we still need one more equation 
from the constraint (\ref{constraint}) as
\beq \label{extra eom}
f_{TT}\dot{T}(\partial^i \varphi
+\frac{1}{2} \partial^i\beta)
+H\partial^i \delta f_T=0.
\eeq
This automatically makes  (\ref{0i}) to be equal 
to (\ref{i0}), which shows the consistency to the 
matter source with $\Theta_{0i}=\Theta_{i0}$.
Although both $\alpha_m$ and $\beta$ are involved
in Eq. (\ref{extra eom}), the contribution of
$\beta$ is in fact separable. Since 
$\partial^2\beta=0$, by taking the gradient
of (\ref{extra eom}) we get
$f_{TT}\dot{T}\varphi=-H \delta f_T$.
Hence, we have three equations (\ref{00}),
(\ref{psi=phi}) and (\ref{extra eom}) to
illustrate the variable $\psi$ in terms of 
$\delta\rho_m$.

In order to simplify the density perturbation equations on sub-horizon 
scales, we also use the quasi-static approximations for those perturbed 
equations (\ref{00})-(\ref{ij}). To be more accuracy, these 
approximations are corresponding to \cite{fR Rev}:
\beq
\frac{k^2}{a^2}\vert X\vert\gg H^2\vert X\vert\,;
\;\;\vert\dot{X}\vert\lesssim \vert HX\vert,
\eeq
where $X=\psi,\varphi,\alpha$ and $\beta$.
It is explicit that, under these approximations, 
Eq. (\ref{extra eom}) indicates nothing but
\beq
\frac{k^2}{a^2}\alpha_m\simeq 0.
\eeq
Consequently, we find from (\ref{00}) with the 
substitution of (\ref{psi=phi}) that
\beq
(1+f_T)\frac{k^2}{a^2}\psi\simeq
\frac{\kappa^2}{2}\delta\rho_m.
\eeq
This implies from (\ref{psi eq}) and (\ref{G eff}) that the
effective gravitational constant is given by
\beq \label{geff}
G_{eff}\simeq \frac{1}{1+f_T}G,
\eeq
where in the TEGR limit of $f_T=const.$, $i.e.$ $f(T)$ being a linear function of $T$, we can
see that the evolution of $\delta_m$  follows that of 
general relativity.
The effective gravitational constant (\ref{geff}) 
is identical to the results obtained 
from the purely metrical approach~\cite{CP of fT,fT p}.

\subsubsection{the growth index}

\begin{figure}[h]
\begin{center}
\includegraphics[width=10 cm]{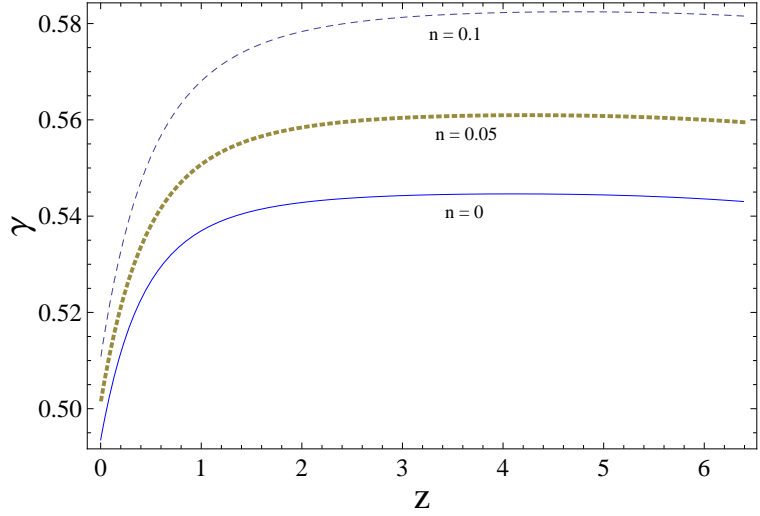}
\end{center}
\caption{Growth index ($\gamma$) as a function of the redshift ($z$) in the power law gravity of $f(T)=\lambda(-T)^n$ with $\Omega^0_m=0.28$,
where the solid, dotted and dashed curves represent $n=0, 0.05$ and $0.1$, respectively.
 }
%\mbox{\epsfig{figure=total2.eps,width=8.74cm,angle=0}} 
\label{g}
\end{figure}

It is straightforward to exam the evolution of 
the matter density perturbation $\delta_m$ 
following the results from the scalar perturbations.
The growth of these small perturbations can
provide structures discriminated from general relativity.
We adopt the parametrization in~\cite{Linder}
with the growth index $\gamma$ given by
\beq
G(a)=\Omega_m(a)^\gamma-1,
\eeq
where $G\equiv d\ln(\delta/a)/d\ln a$.
The matter dominance epoch gives $G(a\ll 1)=0$.
This approach suffices for the analysis beyond
general relativity, provided that the effective gravitational
constant is obtained. We can see that
the perturbation equation (\ref{G eff}) becomes
\beq \label{G}
\frac{d~G}{d\ln a}+\left( 2+\frac{\dot{H}}{H^2}\right)(G+1)+(G+1)^2
= \frac{3}{2}Q\,\Omega_m,
\eeq
where $Q\equiv G_{eff}/G=1/(1+f_T)$ with the effective gravitational
constant given by Eq. (\ref{geff}). The quantity
$Q$ illustrates the deviation from the standard
general relativistic case with $Q=1$ indicating
the $\Lambda$CDM limit. 
As a practical simple example, we consider 
 the power law form of $f(T)$ gravity~\cite{power law}:
\beq \label{power law}
f(T)=\lambda(-T)^n=\lambda(6H^2)^n,
\eeq
where $\lambda=(6H^2_0)^{1-n}(1-\Omega^0_m)/(2n-1)$
is given by the present Hubble parameter
$H_0$ and the present matter density parameter
$\Omega^0_m$.
The condition $n\ll 1$ is required to fit the
current observational data.
It is convenient to define $h\equiv H/H_0$
so that we have
\beq
\frac{\dot{H}}{H^2}=
-\frac{3}{2}\frac{1+f/6H^2+2f_T}{1+f_T-12H^2f_{TT}}=
-\frac{3}{2}\frac{1-h^{2n-2}(1-\Omega^0_m)}{1-nh^{2n-2}(1-\Omega^0_m)}.
\eeq
Given that the evolution of $h$ with $h(z=0)=1$ is
given by
\beq
\frac{dh^2}{d \ln a}=
\frac{-3h^2+3h^{2n}(1-\Omega^0_m)}{1-nh^{2n-2}(1-\Omega^0_m)},
\eeq
we are able to solve the evolution of $G$ numerically from
(\ref{G}) by adopting some fixed values of $n$ and $\Omega^0_m$.
Note that the $G^2$ term in (\ref{G}) is neglected since
the magnitude of today's $G$ is found of order
$-1/2$ even for $n=0.1$~\cite{fT p}. The result in
Fig. \ref{g} reveals the comparison with 
$\Lambda$CDM model ($n=0$) which has the asymptotic
growth index $\gamma_\infty\simeq 0.5454$ when
$z\rightarrow\infty$~\cite{Linder,CDM}.

%=====================================================================================%

%=====================================================================================%
%----------Subsection C. Vector Perturbations------------------------------------%
\subsection{Vector Perturbations}

It is also noteworthy to examine the behavior of the
two unfamiliar vector modes $\alpha^i$ and
$\beta^i$ during the cosmological evolutions.
In the Longitudinal gauge, the torsion tensors from 
(\ref{gauge-ready vector mode}) are given by
\beqa
T^{0}{}_{0i}&=&a\partial_0(G_i+\alpha_i), \nonumber\\ \nonumber
T^{0}{}_{ij}&=&a\left[ \partial_i(G_j+\alpha_j)-\partial_j(G_i+\alpha_i)\right] ,  \nonumber\\
T^{i}{}_{0j}&=&H\delta^i_j+\partial_0 B^i{}_j -a^{-1}\partial_j\alpha^i , \nonumber \\
T^{i}{}_{jk}&=& \partial_j B^i{}_k-\partial_k B^i{}_j.
\eeqa
Given that the scalar functions $\delta T$, $f$ and $f_T$ 
are merely background quantities
in the vector vierbein perturbations,
the $\Theta_i{}^0$ equation of (\ref{eom2}) then directly
yields
\beq \label{v mode}
\frac{1}{2}(1+f_T)\partial^2G_i=a\kappa^{-2}\rho_m\delta u_i^V.
\eeq
Similar to metric perturbations, 
the energy momentum conservation 
$\nabla^\mu \Theta_{\mu\nu}=0$
indicates that the evolution of $\delta u_i^V$ decays as 
$1/a^3$~\cite{Weinberg}, while
  (\ref{v mode})
implies that the vector mode $G_i$ behaves  as $1/a^2$.
Meanwhile, the constraints (\ref{constraint}) provide
\beqa
                   3HG^i&=&a^{-1} \beta^i, \\
\partial^i(G^j+\alpha^j)&=&\partial^j(G^i+\alpha^i),
\eeqa
and it follows that in the Fourier space
the two vector modes
$\alpha^i$ and $\beta^i$ are proportional to
$G^i$, and thus will also decay as $1/a^2$.

We may now give a brief remark on the behavior of 
those 6 EDoFs: $\alpha$, $\beta$, 
$\alpha^i$ and $\beta^i$ in the cosmological perturbations
of $f(T)$ gravity.
It is obvious in the TEGR limit that we 
have the familiar equation of $\psi=\phi$.
Hence, we find from (\ref{psi=phi}) and (\ref{extra eom}) 
that $\alpha$ and $\beta$ are
two new modes in $f(T)$ gravity when comparing
with metric perturbations, while, on the other hand,
$\alpha^i$ and $\beta^i$ are determined by $G^i$ which
give no significance in the cosmological evolution.
In addition to the transverse traceless tensor mode $h_{ij}$,
which is beyond our main concern in the present work,
the vierbein perturbation virtually provides three dynamical 
scalar modes $\psi$(or $\phi$), $\alpha$ and $\beta$.  
This result exactly matches the number of dynamical
degrees of freedom found in~\cite{dof}.

%=====================================================================================%
%----------Subsection C. Vector Perturbations------------------------------------%
\subsection{Super-horizon Scales}

Although the effect of the EDoFs  seems
implicit in  sub-horizon scales,
in the large-scale limit of $k\ll aH$, the
cosmological evolution of the perturbed variables
can be much different. In general relativity, 
one can obtain a simple
solution $\psi=\varphi=const.$ in super-horizon scales, 
which implies the matter perturbation 
$\delta\simeq const.$ \cite{DE}.
For convenience, we denote a new variable
 $\Phi\equiv (\dot{\varphi}+H\psi)/H$ here, 
so that we have $\delta T\simeq 12H^2\Phi$ in $k\ll aH$.
  (\ref{00}), (\ref{0i}) and (\ref{extra eom})
are given by
\beqa \label{s1}
6H^2\left( 1+f_T-12H^2f_{TT}\right)\Phi &\simeq &-\kappa^2\rho_m\delta_m, \\
      \label{s2}
-2H\left( 1+f_T-12H^2f_{TT}\right)\Phi  &\simeq &\kappa^2\rho_m\delta u , \\
      \label{s3}      \dot{H}\varphi=H^2 \Phi, &&
\eeqa
respectively,
while  (\ref{psi eq}) becomes
\beq \label{s4}
\ddot{\delta}_m+2H\dot{\delta}_m \simeq 3\ddot{\Psi}+6H\dot{\Psi},
\eeq
where $\Psi\equiv \psi-H\delta u$.
It is easy to obtain $3H\delta u =\delta_m$ by
comparing  (\ref{s1}) with  (\ref{s2}).
Consequently,  (\ref{s4}) indicates
$\ddot{\psi}+2H\dot{\psi}=0$.
Therefore, we find a simple solution
$\psi=\psi_s=const.$  similar to the case 
in general relativity.

Substituting (\ref{s3}) into  (\ref{s1})
and using the background relations (\ref{bg}),
we finally arrive at
\beq \label{s5}
\delta_m\simeq 3\varphi=
3\psi_s -36\frac{\dot{H}f_{TT}}{1+f_T}\alpha_m,
\eeq
where  (\ref{psi=phi}) has  also been used.
It is evident that in the TEGR limit we obtain the
expected result $\delta_m\simeq const$. 
The last term in  (\ref{s5}) addresses the
deviation from general relativity as the 
impact of the new degree of freedom,
$\alpha_m$. 
It has been demonstrated numerically that the 
deviation in $f(T)$ gravity
 from $\Lambda CDM$ becomes inevitably
apparent at some rather larger scale 
$k\sim 10^{-4}\,h\, Mpc^{-1}$ \cite{LSS}.

%==================================================================================================================================%
%----------------------------------------Section IV. Conclusions--------------------------------------------------------------------%
\section{Conclusions}

We have investigated the
matter density perturbations for modified teleparallel
gravity models of dark energy. The lack of local
Lorentz symmetry in the teleparallel formulation
introduces dynamical degrees of freedom beyond
the metric scenario and hence, a complete formula
for vierbein perturbations is required.
We have found from a specific decomposition of the
vierbein field that the 6 extra degrees of freedom
can be illustrated by two scalar and
two transverse vector modes. 

We have also applied the vierbein perturbation in $f(T)$
gravity to examine particularly the cosmological implication 
of those unfamiliar variables, 
i.e. $\alpha$, $\beta$, $\alpha^i$ and $\beta^i$.
Our study has been simplified by imposing
the perfect fluid assumption to the matter source
on both background and perturbation levels.
Although the result in sub-horizon scales indicates
no significant contribution from the EDoFs, 
we have shown that the density perturbation in
super-horizon scales is indeed affected by the
new scale modes.

In summary, given that the two transverse vectors
 $\alpha^i$ and $\beta^i$ are mere decaying modes,
we have matched the number of physical degrees of
freedom found in modified teleparallel theories.
Namely, the three degrees of freedom other than
the transverse-traceless tensor are addressed by
one usual scalar plus the two new scalar modes of
$\alpha$ and $\beta$. Nevertheless, it remains
interesting that only the scalar $\alpha$ tends
to show up the physical importance in the 
present study, while the mode $\beta$ involves
in the results nowhere. The significance of
such a scalar mode could be worthy of
the further 
investigation in teleparallel gravity theories.

%==================================================================================================================================%
%----------------------------------------Acknowledgments--------------------------------------------------------------------%
%\noindent
\begin{acknowledgments}
This work was partially supported by National Center of Theoretical
Science and  National Science Council (NSC-98-2112-M-007-008-MY3 and
NSC-101-2112-M-007-006-MY3) of R.O.C.
\end{acknowledgments}

%==================================================================================================================================%
%----------------------------------------Bibliograpy--------------------------------------------------------------------%

\end{document}